\documentclass{PoS}
\usepackage{graphicx}
\usepackage[subfigure]{graphfig}
\usepackage{amsmath}
\usepackage{epsfig}
\usepackage{epstopdf}

\def\be{\begin{equation}}
\def\ee{\end{equation}}
\def\bea{\begin{eqnarray}}
\def\eea{\end{eqnarray}}

\definecolor{green}{rgb}{0,.5,0}

\title{A Lattice Story of Proton Spin}

\ShortTitle{Proton Spin Decomposition}

\author{\speaker{Yi-Bo Yang}\\
       Institute of Theoretical Physics, Chinese Academy of Science, Beijing, 100190, China\\
       E-mail: \email{ybyang@itp.ac.cn}
\vspace*{-0.5cm}
\begin{center}
\large{
\vspace*{0.4cm}
\includegraphics[scale=0.20]{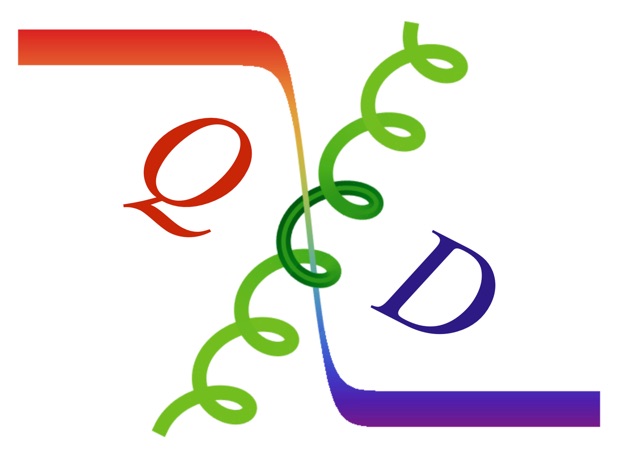}\\
\vspace*{0.4cm}
($\chi$QCD Collaboration)
}
\end{center}
}

\abstract{In this contribution, I summarized the recent Lattice QCD consensuses on the quark helicity, plus the investigations on the gluon helicity and orbital angular momenta.  The preliminary non-perturbative normalized and renormalized Ji quark and gluon angular momentum results are also reported and compared with the previous 1-loop perturbative renormalized results.}

\FullConference{36th International Symposium on Lattice Field Theory - LATTICE 2018\\
        July 22 - July 38, 2018\\
        Michigan State University, East Lansing, Michigan, USA}

\begin{document}

\section{Introduction}

The old-fashion understanding of the proton spin in view of the quark spin is very simple: the proton includes three quarks with spin 1/2, thus the proton spin can be natively 1/2 based on the restriction of the angular momentum coupling. In the SU(6) quark model, one can count the number of quarks in the proton with certain helicity as shown in the introduction of Ref.~\cite{Zhang:2012sta}:
\bea
u^{\uparrow}=5/3,\ u^{\downarrow}=1/3,\ d^{\uparrow}=1/3,\ d^{\downarrow}=2/3,\ \textrm{and}\ s^{\uparrow}=s^{\downarrow}=0.
\eea
Thus one can have the naive sum rule $u^{\uparrow}+u^{\downarrow}=2$, $d^{\uparrow}+d^{\downarrow}=1$, and $\frac{1}{2}\sum_{q=u,d,s} \Delta q=\frac{1}{2}$  satisfied (where $\Delta q\equiv q^{\uparrow}-q^{\downarrow}$), and predict the iso-vector quark spin in the proton $g_A^3\equiv\Delta u-\Delta d$ to be 5/3. 

But if the iso-spin breaking effect between the $u$ and $d$ quark (and also the possible electric-weak effect) is neglect, $g_A^3$ can be related to the width of the neutron weak decay based on the iso-spin symmetry of the operator $\bar{u}u-\bar{d}d$ and $\bar{u}d$, and the experiment shows that $g_A^3=1.2723(23)$~\cite{Patrignani:2016xqp}.  Even more, $\Delta q$ can be obtained through the integration of the polarized parton distribution function 
\bea
\Delta q(x)=\int \frac{\textrm{d}\xi^{-}}{2\pi} e^{-ixP^{+}\xi^{-}}\langle PS|\bar{q}(\xi^-)\gamma_5\gamma^{+}e^{\int^{\xi^-}_0 igA_{+}(\eta)\textrm{d}{\eta} }q(0)|PS\rangle,
\eea
which can be determined based on the global fit of the deep inelastic scattering experiments. Based on the recent COMPASS fit~\cite{Adolph:2015saz}, 
\bea
\Delta u\in [0.82,0.85],\ \Delta d\in[-0.45, -0.42],\ \textrm{and}\ \Delta s\in [-0.11, -0.08],
\eea 
at $Q^2$=3 GeV$^2$, and such a result is consistent with the $g^3_A$ result from the neutron weak decay, but obviously different from the prediction of the SU(6) quark model. The total helicity contribution from quark, is just about 30\% of the proton spin~\cite{deFlorian:2009vb,Nocera:2014gqa,Adolph:2015saz}, and the missing parts should come from the gluon helicity (which is known to be around 40\% at $Q^2$=10 GeV$^2$~\cite{deFlorian:2014yva}) and also the orbital angular momenta of quark and gluon.

Since the quarks are relativistic and the strong interaction exists between quark and gluon, the failure of the SU(6) quark model is expectable and then an better and even accurate theoretical prediction of the proton spin components is desired and becomes a challenge to our understanding on QCD. As the native tool to calculate the non-perturbative characters of QCD, Lattice QCD takes the responsibility to repeat the experimental quark helicity results from the first principle, to check the control of the systematic uncertainties before make prediction on the less-known quantities. $\chi$QCD collaboration has spent over 20 years on the lattice QCD calculation of the proton spin components, along with the efforts from the other lattice groups. In the recent lattice calculations, some consensuses have been reached in the quark helicity, and several investigations are made for the gluon helicity and also orbital angular momenta.  In this contribution, I will briefly review those progress and provide an outlook on possible further studies.

\section{Quark helicity}

\begin{figure}[htb]
  \includegraphics[width=1.0\hsize,angle=0]{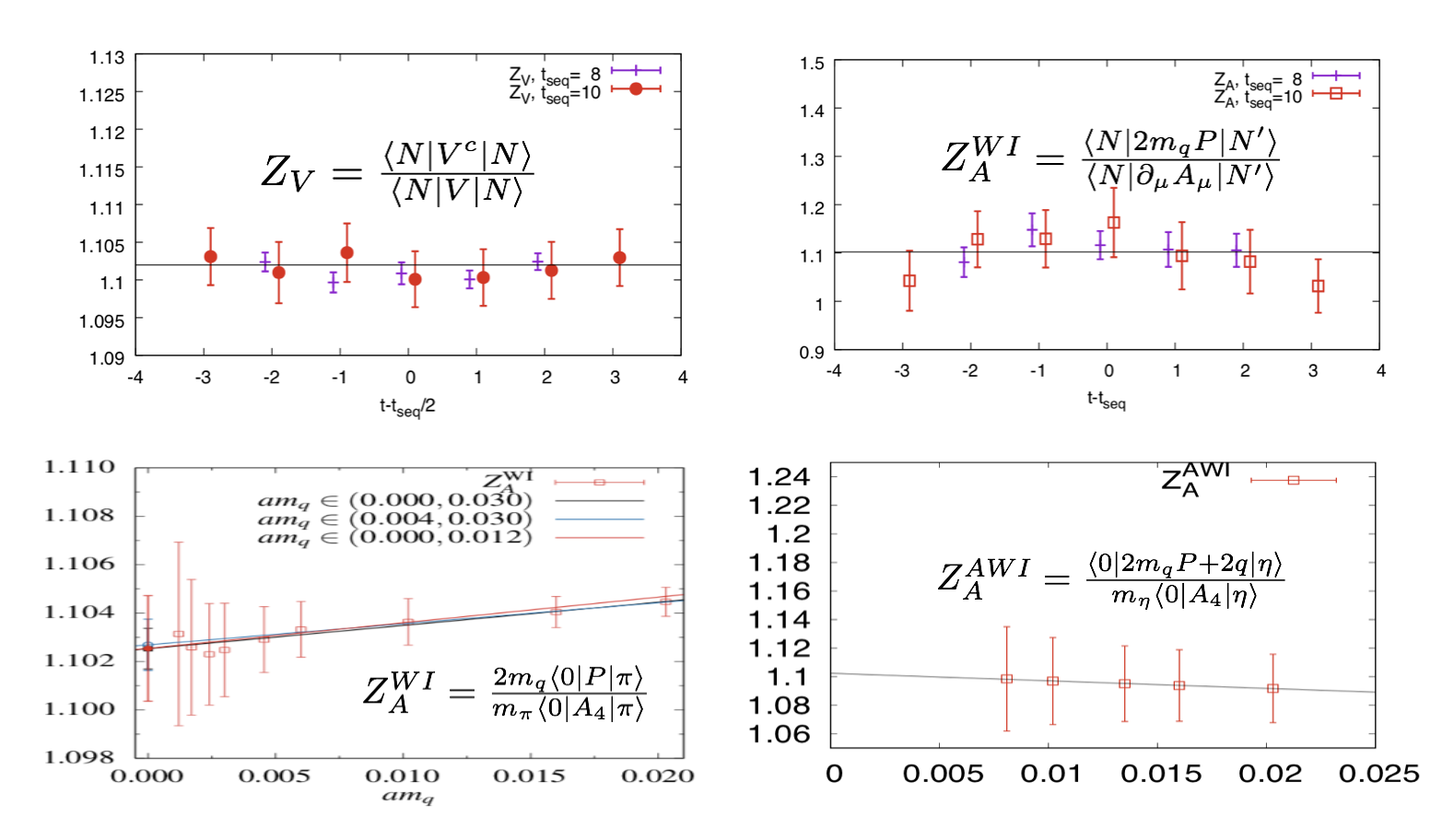}
\caption{The normalization constant of the vector and axial-vector charge from the (anomalous) Ward Identity. The left-top panel shows the ratio of the vector charge in the nucleon with the conserved vector current and local one. The ratio of the nucleon matrix element with the right and left hand sides of the axial-vector Ward Identity without anomaly is presented in the right-top panel. These two ratios define the vector and axial-vector normalization constant through the nucleon matrix element, and the consistency of them shows the chiral symmetry conserved well using the overlap fermion, without additional breaking due to the lattice artifacts. The lower two panels show the axial-vector normalization w.o. and with the triangle anomaly defined through the pseudo-scalar meson amplitude (left-bottom and right-bottom panels). They agree with each other and also those defined from the nucleon matrix elements.} \label{fig:Zwi}
\end{figure}

The Lattice QCD calculation of the quark helicity in proton is straight forward, 
\bea
\Delta q=\langle PS|\bar{q}\gamma_5\vec{\gamma}\cdot\vec{S}q|PS\rangle=\frac{\langle \Gamma^m_3\int \textrm{d}^3y\chi(\vec{y},t_f)\int \textrm{d}^3 x{\cal A}^q_3(\vec{x},t)\bar{\chi}(\vec{0},0)\rangle}{\langle \Gamma^e\int \textrm{d}^3y\chi(\vec{y},t_f)\bar{\chi}(\vec{0},0)\rangle}|_{t\rightarrow \infty,t_f-t\rightarrow \infty},
\eea
where $|PS\rangle$ is the nucleon state with momentum $P$ and polarization $S$, $\chi$ is the nucleon interpolation field, ${\cal A}_{\mu}=\bar{\psi}\gamma_5\gamma_{\mu} \psi$ is the axial-vector current, and $\Gamma^e$ and $\Gamma^m_3$ are the unpolarized projection operator of the proton and the polarized one along the z-direction, respectively.

But the practical calculation suffers from kinds of the systematic uncertainties. Besides the chiral/continuum/volume extrapolations which exist in all kinds of the Lattice QCD calculation, the $\Delta q$ calculation also requires proper treatments on the normalization/renormalization issue and also the excited-state contamination (as we have to extrapolate $t$ and $t_f-t$ to $\infty$). 

In the continuum, the axial-vector current is free of the renormalization in the non-singlet cases, which is guaranteed by the Ward Identity. But under the lattice regularization, both the axial-vector and vector current suffer from additional normalization, and such a normalization can be different in these two currents due to the additional chiral symmetry breaking unless the chiral fermion is used. Fig.~\ref{fig:Zwi} shows four ways to obtain this normalization.  The calculation with Overlap fermion satisfying $\{D_c, \gamma_5\} = 0$~\cite{Chiu:1998gp} on the HYP smeared RBC 24I ensemble (Domain wall sea and Iwasaki gauge action)~\cite{Blum:2014tka} shows that the normalization constant from kinds implementations (the ratio of the conserved/local vector nucleon matrix element~\cite{Hasenfratz:2002rp}, right/left hand side of the axial-vector Ward Identity in the nucleon matrix element and pseudo-scalar meson amplitude) are all consistent with each other, as the chiral symmetry is conserved.

\begin{figure}[htb]
 \centerline{ \includegraphics[width=0.8\hsize,angle=0]{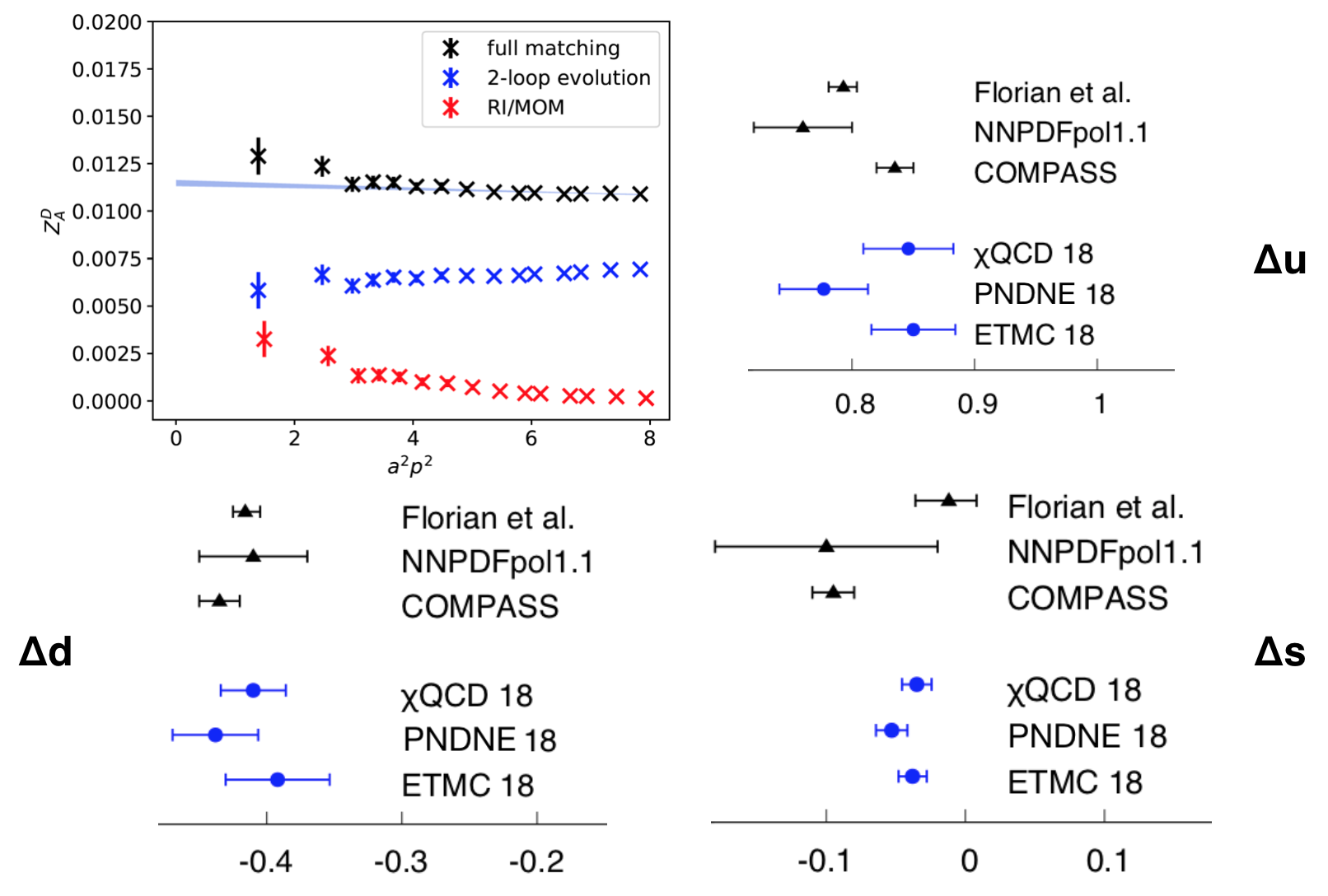} }
\caption{The 2-loop renormalization effect of the quark spin (left-top panel)~\cite{Liang:2018pis} and the summary plots of $\Delta u/d/s$ from the phenomenology determinations and Lattice QCD calculation. As in the left-top panel, the renormalization effect of the axial-vector current can be 1\% level per flavor, as both the evolution and finite matching effects enhanced the tiny RI/MOM mixing between different flavors~\cite{Liang:2018pis}. The present Lattice QCD results using Overlap fermion on Domain wall sea~\cite{Liang:2018pis}, Clover fermion on HISQ sea~\cite{Lin:2018aa} and unitary Twisted mass (with clover term) fermion~\cite{Alexandrou:2017oeh} agree with each other within two sigma, and also the phenomenology determinations from Florian et.al.~\cite{deFlorian:2009vb}, NNPDFpol1.1~\cite{Nocera:2014gqa} and COMPASS~\cite{Adolph:2015saz}. Even more, the present Lattice QCD prediction of the $\Delta s$ has been more precise than the phenomenology determinations. } \label{fig:quark_spin}
\end{figure}

For the singlet quark helicity, the renormalization at 2-loop level is required due to the anomalous Ward Identity. Since such a renormalization effect under the lattice regularization can only be handled by non-perturbative renormalization use the off-shell RI/MOM scheme, a 2-loop matching is required to connect the RI/MOM renormalization constant to that under the $\overline{\textrm{MS}}$ scheme, besides the scheme evolution~\cite{Green:2017keo}. As in the left-top panel of Fig.~\ref{fig:quark_spin}, the matching effect for the overlap fermion turns out to be comparable with the evolution effect~\cite{Liang:2018pis}. The recent dynamical $\Delta u/d/s$ results are summarized in the rest panels of Fig.~\ref{fig:quark_spin}, and the $n_f\times$1\% renormalization effect mentioned above is still covered by the statistical uncertainty unless the accuracy of $\Delta u/d/s$ can be improved by a factor of 3 or so. Even though the lattice results using three different actions~\cite{Liang:2018pis,Lin:2018aa,Alexandrou:2017oeh} suffer from different uncontrolled systematic uncertainties, they agree with each other within 2 $\sigma$ and can provide more precise prediction on $\Delta s$, comparing to the present phenomenology determinations. The unitary overlap fermion simulation at $a$=0.11 fm with down to 290 MeV pion mass~\cite{Yamanaka:2018uud} also support the same consensus.

\begin{figure}[htb]
  \includegraphics[width=1.0\hsize,angle=0]{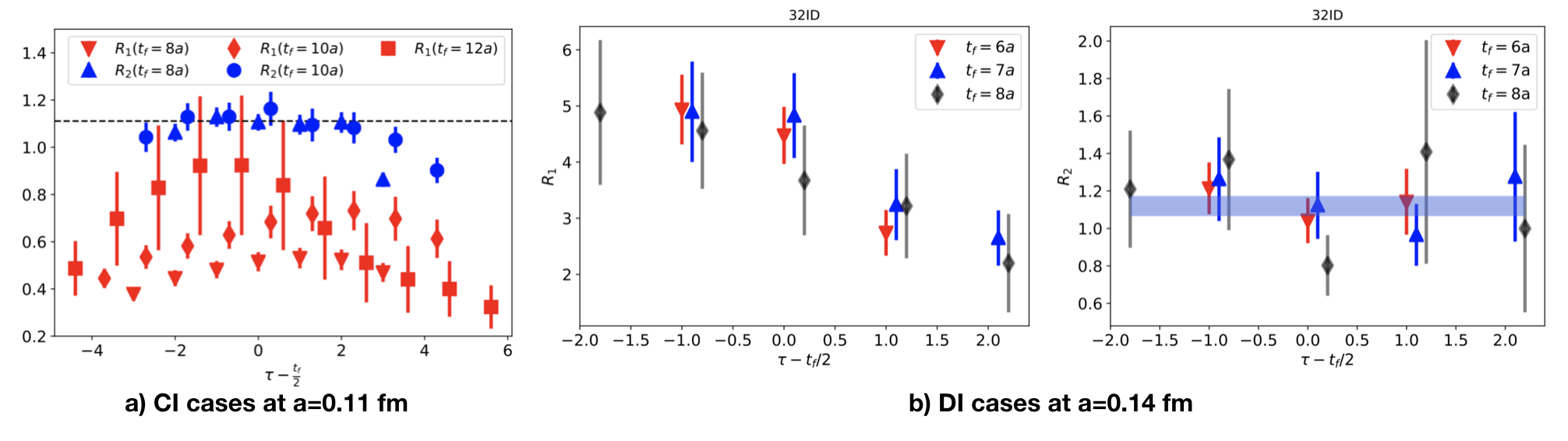}
\caption{The ratio $R_1$ as the ratio of the right and left hand sides of the axial-vector (anomalous) Ward Identity, based on the form factor extracted from the spatial components of the axial-vector current; and $R_2$ which used the discretized definition of the left hand side and then involves the temporal components of the axial-vector current~\cite{Liang:2018pis}. The left panel shows that the connected insertion case (without anomaly) based on the calculation at a=0.111 fm, and the disconnected plots for $R_{1,2}$ at a=0.143 fm are shown in the rest two panels respectively. $R_2$ is insensitive to the source-sink separations and agrees with that obtained in Fig.~\ref{fig:Zwi}, while $R_1$ suffers from huge excited state contamination and can not be fully cured by two-state fit with the data using up to $1.2 fm$ source-sink separation. } \label{fig:excited}
\end{figure}

But whether the excited-state contamination has been properly removed, is still an open question. Let us start from the off-forward condition, where the following ratio can be defined using the axial-vector anomalous Ward Identity,
\bea
R_2&=&\frac{\langle \Gamma^m_3\int \textrm{d}^3y \textrm{d}^3x e^{-\vec{q}\cdot(\vec{y}-\vec{x})}\chi(\vec{y},t_f)\big((2m_q{\cal P}(\vec{x},t)-2i\mathrm{q}(\vec{x},t)\big)\bar{\chi}(\vec{0},0)\rangle}
{\langle \Gamma^m_3\int \textrm{d}^3y \textrm{d}^3x e^{-\vec{q}\cdot(\vec{y}-\vec{x})}\chi(\vec{y},t_f)\partial_{\mu}{\cal A}_{\mu}(\vec{x},t)\bar{\chi}(\vec{0},0)\rangle}\nonumber\\
&=&\frac{\langle \Gamma^m_3\int \textrm{d}^3y \textrm{d}^3x e^{-\vec{q}\cdot(\vec{y}-\vec{x})}\chi(\vec{y},t_f)\big((2m_q{\cal P}(\vec{x},t)-2i\mathrm{q}(\vec{x},t)\big)\bar{\chi}(\vec{0},0)\rangle}
{\langle \Gamma^m_3\int \textrm{d}^3y \textrm{d}^3x e^{-\vec{q}\cdot(\vec{y}-\vec{x})}\chi(\vec{y},t_f)\big(i\vec{q}\cdot\vec{{\cal A}}(\vec{x},t)+{\cal A}_4(\vec{x},t)-{\cal A}_4(\vec{x},t-1)\big)\bar{\chi}(\vec{0},0)\rangle}
\eea
where ${\cal P}=\bar{\psi}\gamma_5\psi$, $\mathrm{q}=\frac{\alpha_s}{4\pi}F\tilde{F}$, and $R_2$ should be independent to $t$ and $t_f$ and equal to the normalization mentioned in the previous part of this section. But we can further replaced the contribution from ${\cal A}_4$ into the combination of ${\cal A}_i$ through the definition of the axial-vector form factors ($q=P'-P$),
\bea
\langle P'S'|{\cal A}_{\mu}|PS\rangle=\bar{\chi}(P',S')[i\gamma_{\mu}\gamma_5g_A(q^2)-iq_\mu\gamma_5h_A(q^2)]\chi(P,S)
\eea
which only holds in the $t,t_f\rightarrow \infty$ limit, to get $R_1$ (see Ref.~\cite{Liang:2018pis} for the details). $R_1$ should equals to $R_2$ in the  $t,t_f\rightarrow \infty$ limit, while can have large breaking at finite $t$ and $t_f$. As in Fig.~\ref{fig:excited}, $R_1$ suffers from obvious excited state contamination in both the connected insertion (left panel) and also the disconnected insertion (middle panel) cases, and can not be cured by two-state fit with the data using up to 1.2 $fm$ source-sink separation. It is also observed in the PNDME calculation~\cite{Rajan:2017lxk}, and hinted by our previous forward matrix element calculation based on both ${\cal A}_i$ and ${\cal A}_4$. Recent $\chi$PT development suggests that it would be a contamination from the $\pi$N state and then can still be huge in the ${\cal A}_4$ case and sizable in ${\cal A}_i$ cases with even 2 fm source-sink separation, while such a contamination can be removed properly~\cite{Bar:2018xyi}.

\begin{figure}[htb]
 \centerline{ \includegraphics[width=0.6\hsize]{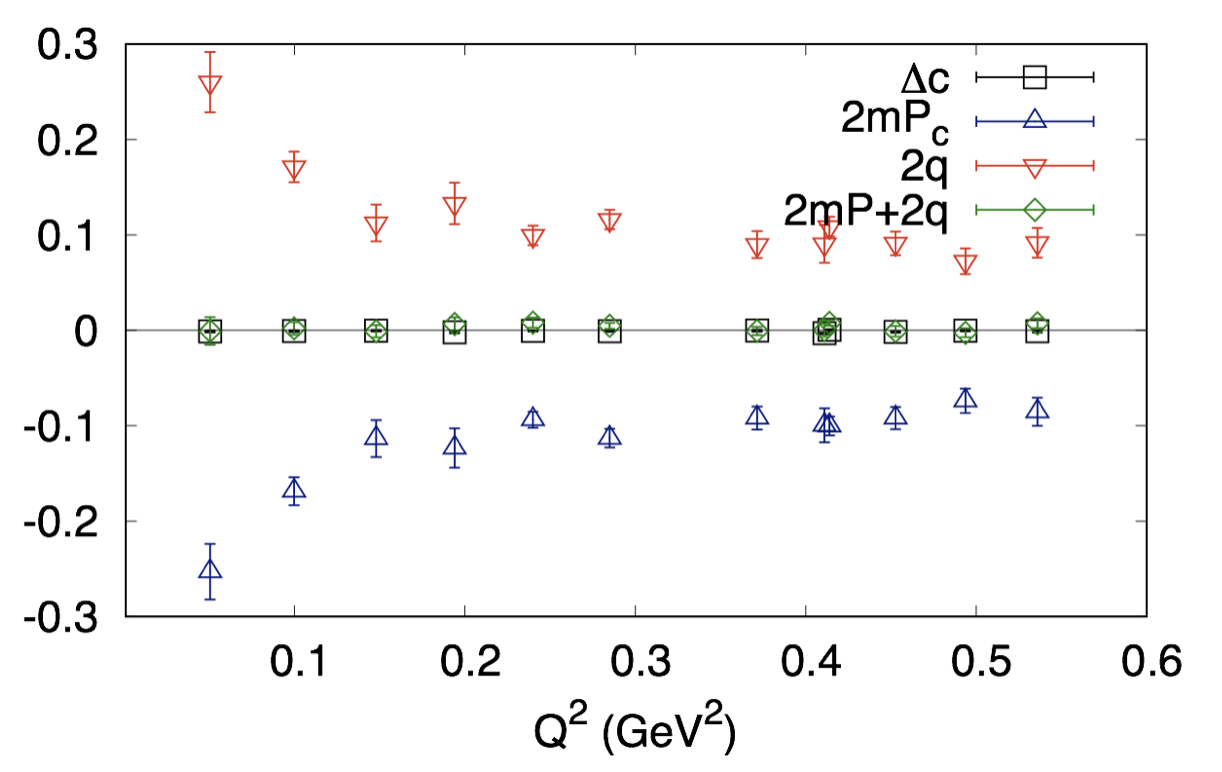}}
\caption{Even though the charm quark helicity $\Delta c$ (black boxes) is small, but the contributions from both the terms of the right hand side of the anomalous Ward Identity, pseudo scalar current (dark blue triangles) and chiral anomaly (red triangles), can be large and cancel each other (green diamonds).} \label{fig:quark_spin_charm}
\end{figure}

Comparing to $\Delta u/d/s$, $\Delta c$ is known to be small from the direct Lattice QCD calculation on the heavy quark bi-linear axial-vector current. But in view of the anomalous Ward Identity, one can further decompose the axial-vector current into two terms, the pseudo scalar and topological charge terms,
\bea
\int d^3x \vec{\cal A}(x) \ =\ 2 m_f  \int d^3x\ \vec{x} P(x) - 2 i \int d^3x\ \vec{x}\ \mathrm{q}(x),
\eea
and calculate these two terms at finite momentum transfer $\vec{q}$ and then extrapolate to the forward limit $Q^2\rightarrow 0$~\cite{Gong:2015iir}. Fig.~\ref{fig:quark_spin_charm} illustrate an update on smaller $Q^2$ comparing to what used in Ref.~\cite{Gong:2015iir}, and the result shows that even though the $\Delta c$ is small, the contribution from both pseudo scalar and topological charge contributions are large and almost the same for the heavy quark case except the sign.

\section{Gluon helicity}

The gluon helicity $\Delta g$ is defined through the integration of the gluon helicity distribution,
\bea\label{eq:glue_helicity}
\Delta g(x)=\int \frac{\textrm{d}\xi^{-}}{2\pi xP^{+}} e^{-ixP^{+}\xi^{-}}\langle PS|F_a^{\mu+}(\xi^-)(e^{\int^{\xi^-}_0 igA_{+}(\eta)\textrm{d}{\eta} })_{ab}\tilde{F}_{\mu,b}^{\ +}(0)|PS\rangle,
\eea
where the light-cone gauge link is defined in the adjoint representation and corresponds to a pair of the gauge links to make the non-local pair of $F$ to be gauge invariant. By integrating over x and take the light-cone gauge, the definition of $\Delta g$ can be simplified into
\bea
\Delta g=\langle PS|\vec{E}^a\times \vec{A}^a \cdot \vec{s} |PS\rangle.
\eea
The phenomenology fit of the experiment has shown that the $\Delta g$ is around 0.4 at $\overline{\textrm{MS}}$ 10 GeV$^2$ with good signal if the contribution of $\Delta g(x)$ from $x<0.05$ is dropped~\cite{deFlorian:2014yva}, but $\Delta g$ is not directly calculable with Lattice QCD as the
 light-cone gauge condition can not be implemented in the euclidean space.
 
  At the same time, the large momentum effective theory (LaMET)~\cite{Hatta:2013gta} suggests that the Coulomb and also temporal gauge fixing condition become to the light-cone one when the nucleon is boosted to the infinite momentum frame (IMF), and then the $\vec{E}\times \vec{A}$ nucleon matrix element under either Coulomb or the temporal gauge, the ``gluon spin'' $S_g$, becomes to the gluon helicity $\Delta g$ in the IMF with proper matching for their difference in the UV behavior.

\begin{figure}[htb]
 \includegraphics[width=0.5\hsize]{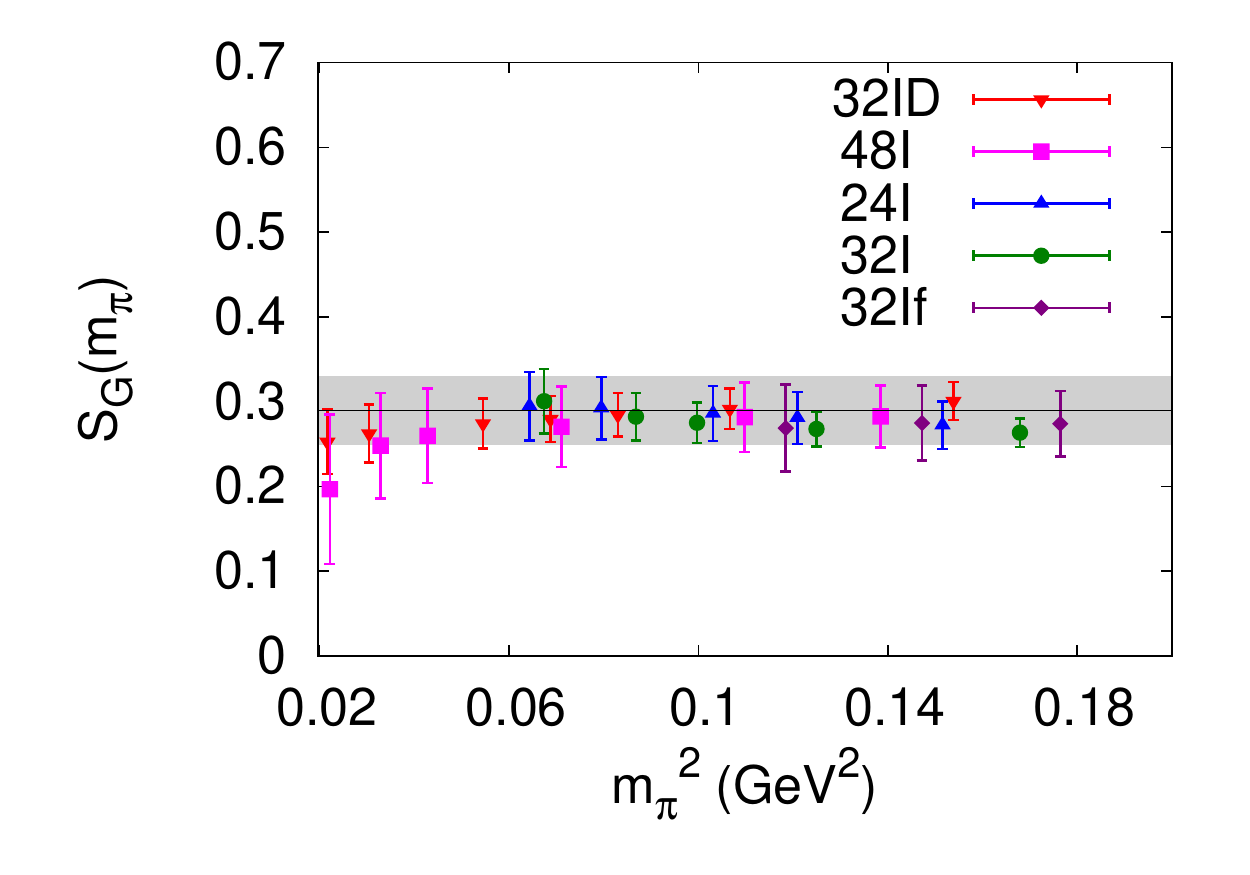}
 \includegraphics[width=0.5\hsize]{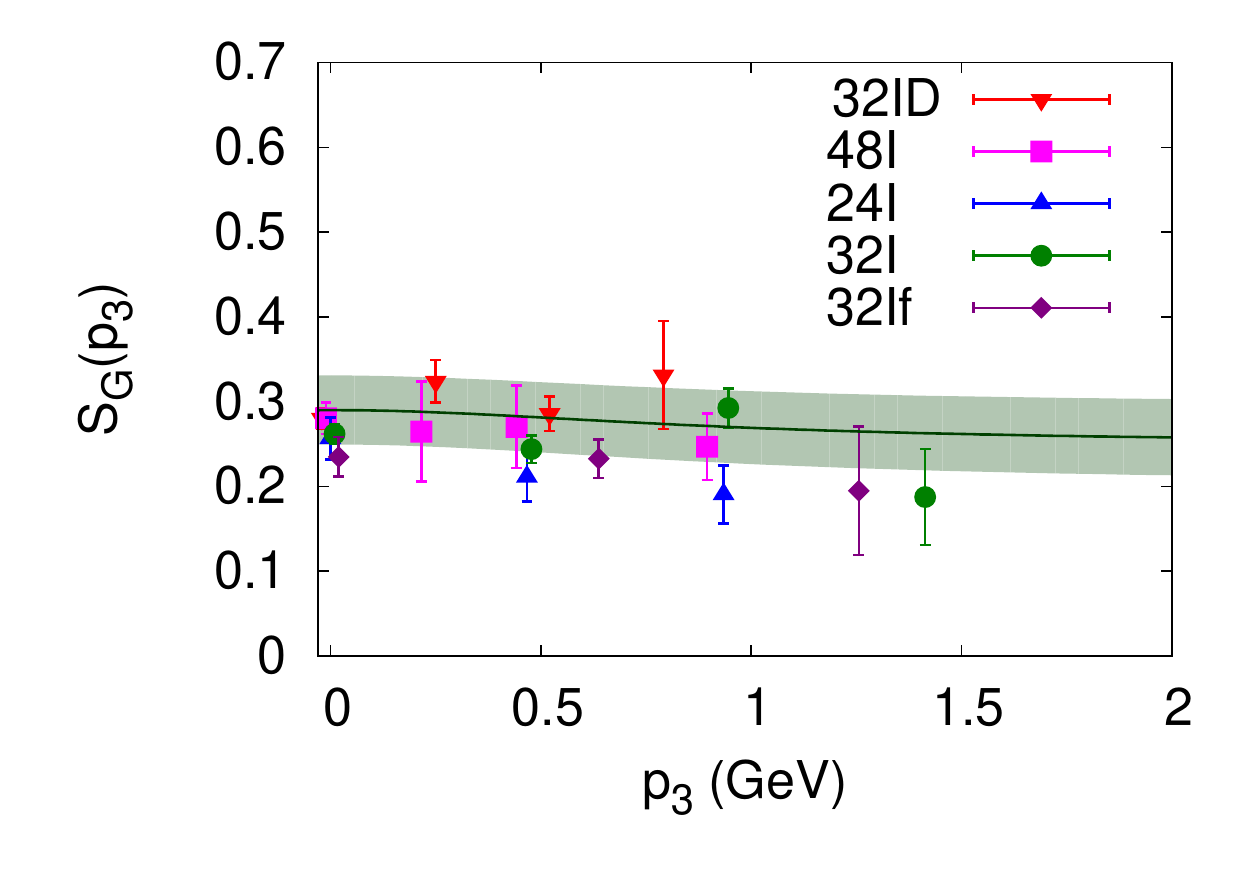}
\caption{The quark mass and momentum dependence of the glue spin $S_g$~\cite{Yang:2016nfc}, as shown in the left and right panel respectively. Both the dependence are mild. The result has good signal in the rest frame while those in the moving frame should be improved with the momentum smearing technique~\cite{Bali:2016lva} before the prediction on the $\Delta g$ can be made.} \label{fig:glue_spin}
\end{figure}

The Coulomb gauge case, which is equivalent to the gluon spin defined by X. Chen et.al.~\cite{Chen:2009mr,Zhao:2015kca}, are studied with Lattice QCD on 4 kinds of the lattice spacing in the pion mass range [140, 400] MeV and up to nucleon momentum $P_z=1.4$ GeV~\cite{Yang:2016nfc}. As in Fig.~\ref{fig:glue_spin}, the result is insensitive to the pion mass and also the investigated nucleon momentum, while the statistical precision with large $P_z$ should be improved with the momentum smearing technique~\cite{Bali:2016lva} in the future. To reach the prediction on the $\Delta g$, the LaMET matching should be obtained at 2-loop level since the 1-loop correction turns out to be comparable with the tree level value. 

\begin{figure}[htb]
 \centerline{ \includegraphics[width=0.6\hsize]{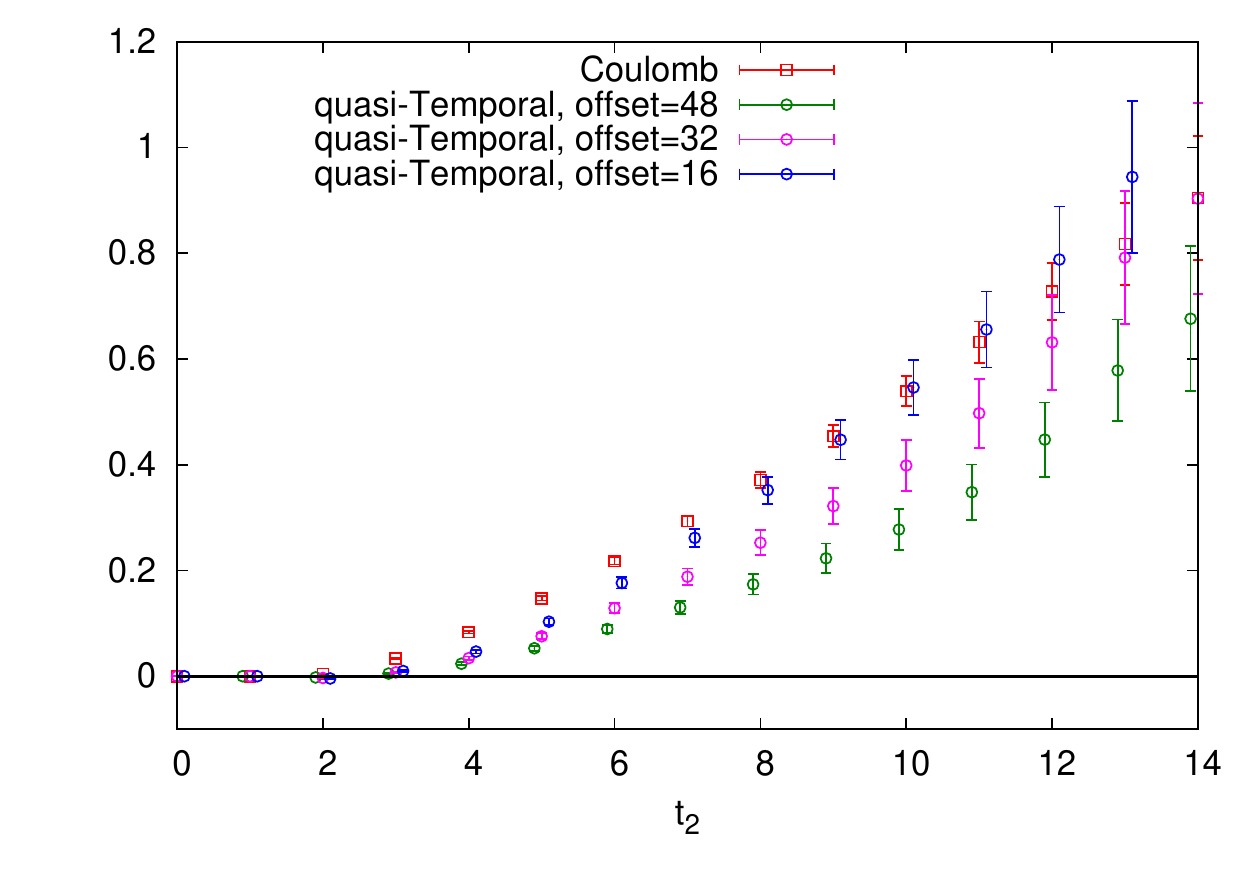}}
\caption{The preliminary result of the glue spin under the quasi-temporal gauge. The figure shows the summed matrix elements as the function of the source-sink separation $t_2$, with the reference Coulomb gauge fixed time slide is 0, 16, 32 and 48 time slide away from the time slide of the current operator $E\times A$ (the red boxes and blue/pink/green dots, the result in the first case is exactly the same as that under the coulomb gauge). As in the figure, the slope at large $t_2$ which corresponds to the ground state matrix element, are insensitive to the place of the reference time slide, while the excited state contamination become larger with larger current-reference separation.} \label{fig:glue_spin_temporal}
\end{figure}

In the other hand, the calculation of the gluon spin under the temporal gauge fixing condition $A_0=0$ would not require any matching at 1-loop level to reach $\Delta g$~\cite{Hatta:2013gta}, while $A_0=0$ can not be implemented on a finite size lattice since the Polyakov loop is non-zero and gauge invariant. There are two alternative solutions: 

1. The quasi temporal gauge with the condition $\partial_iA_i(\vec{x},t_0)=0, A_0(\vec{x},t)=0$ for all the $x$ and $t\neq t_0$~\cite{Conti:1995dw}. Such a condition is doable while the translation invariance is broken. Fig.~\ref{fig:glue_spin_temporal} shows the preliminary summed ratio
\bea
\sum_\tau\frac{\langle \Gamma^m_3\int \textrm{d}^3y\chi(\vec{y},yt_2)(\int \textrm{d}^3x \vec{E}\big(\vec{x},\tau)\times \vec{A}(\vec{x},\tau)\big)_3\bar{\chi}(\vec{0},0)\rangle}{\langle \Gamma^e\int \textrm{d}^3y\chi(\vec{y},yt_2)\bar{\chi}(\vec{0},0)\rangle} \ _{\overrightarrow{t_2\rightarrow \infty}} t_2 S^{qt}_g+{\cal O}(1)
\eea
 on the RBC 24I ensemble~\cite{Blum:2014tka} with $a$=0.111 fm and  $P_z$=0, with different separation between the time slide $\tau$ and the reference time slide $t_0$ satisfying $\partial_iA_i(\vec{x},t_0)=0$. The slope at large $t_2$ corresponds to the gluon spin under the quasi-temporal gauge $S^{qt}_g$, which is insensitive to the separation $|\tau-t_0|$ even though the  the excited state contamination become larger with larger $|\tau-t_0|$. Note that the $|\tau-t_0|=0$ case is 
 exactly the same as the case under the Coulomb gauge.
 
 2. The Polyakov gauge with the condition $\partial_0 A_0=0$. In contrast with the temporal gauge, the Polyakov gauge condition doesn't break the translation invariance or the gauge invariance of the Polyakov loop. The related matching to $\Delta g$ would be similar to the temporal gauge case while should be investigated. But the preliminary attempt shows that the $S^{P}_g$ under the Polyakov gauge is consistent with zero in the rest frame, a non-zero $P_z$ or residual gauge fixing would be required to see the non-zero result.
  
All the above $S_g$'s depend on the gauge fixing condition and such a dependence can only be removed at IMF with perturbative LaMET matching. LaMET provides another possibility to construct a gauge invariant quasi-gluon helicity, through the quasi-PDF approach,
\bea
\Delta \tilde{g}(x)=\int_{0}^{\infty} \textrm{d}z \Delta \tilde{H}_g(z)|_{P_z\rightarrow \infty}=\Delta g(x)+{\cal O}(\alpha_s).
\eea
where
\bea
\Delta \tilde{H}_g(z)=\sum_{i=x,y}\langle PS|F_{iz,a}(z)(e^{\int^{z}_0 igA_z(z')\textrm{d}{z'} })_{ab}\tilde{F}_{iz,b}(0)|PS\rangle
\eea
and the ${\cal O}(\alpha_s)$ correction can be eliminated by matching the the matrix elements $\Delta \tilde{H}_g(z)$ perturbatively to their light-cone counter parts in Eq.~\ref{eq:glue_helicity} in the coordinate space first before the integration. But the this approach suffers from some cutoff uncertainty: It has to drop the contribution from very large $z$, likes the small $x$ contribution of $\Delta g(x)$ will be dropped in the phenomenology determination of $\Delta g(x)$. 

\section{Orbital angular momenta}

The rest part of the proton spin, should come from the orbital angular momenta (OAM) of both the quark and gluon. There are two popular definitions of the complete proton spin decomposition: One is proposed by Jaffe and Manohar, from the QCD energy energy momentum tensor (EMT) without the symmetrization, and can be connected to the collider experiments through the Wilson links along the light-cone direction~\cite{Jaffe:1989jz}. It equivalent to the following definition under the light-cone gauge fixing condition $A_{LC}^{+}=0$,
\bea
J=\frac{1}{2}\bar\psi\,\vec{\gamma}\,\gamma^5 \,\psi+ \psi^\dag \,\{ \vec{x} \times (i \overrightarrow{\partial}_{LC}) \} \,\psi+ \vec{E}_{LC,a}\times \vec{A}_{LC,a}+E^{i}_{LC,a} \vec{x}\times \overrightarrow{\partial}_{LC} A^{i}_{LC,a},
\eea
where four terms in the right hand side correspond to quark helicity, JM OAM, gluon helicity and JM gluon OAM respectively. The other one is proposed by Ji, based on the symmetrized QCD EMT, gauge invariant and frame independent~\cite{Ji:1996ek},
\bea
J=\frac{1}{2}\bar\psi\,\vec{\gamma}\,\gamma^5 \,\psi+ \psi^\dag \,\{ \vec{x} \times (i \overrightarrow{D}) \} \,\psi+ \vec{x}\times (\vec{E}_{a}\times \vec{B}_{a}),
\eea
where the second and third terms in the right hand side are Ji quark OAM and Ji gluon total angular momentum (AM). The quark spin and Ji can be combined into Ji quark AM with a simple form $\frac{1}{4}\bar{\psi} \,\{ \vec{x} \times (i \overleftrightarrow{D}_{\{4}\gamma_{i\}}) \} \,\psi$. Note that the Ji quark and gluon AM correspond to the first moments of their GPD.

\begin{figure}[htb]
 \centerline{ \includegraphics[width=1.0\hsize]{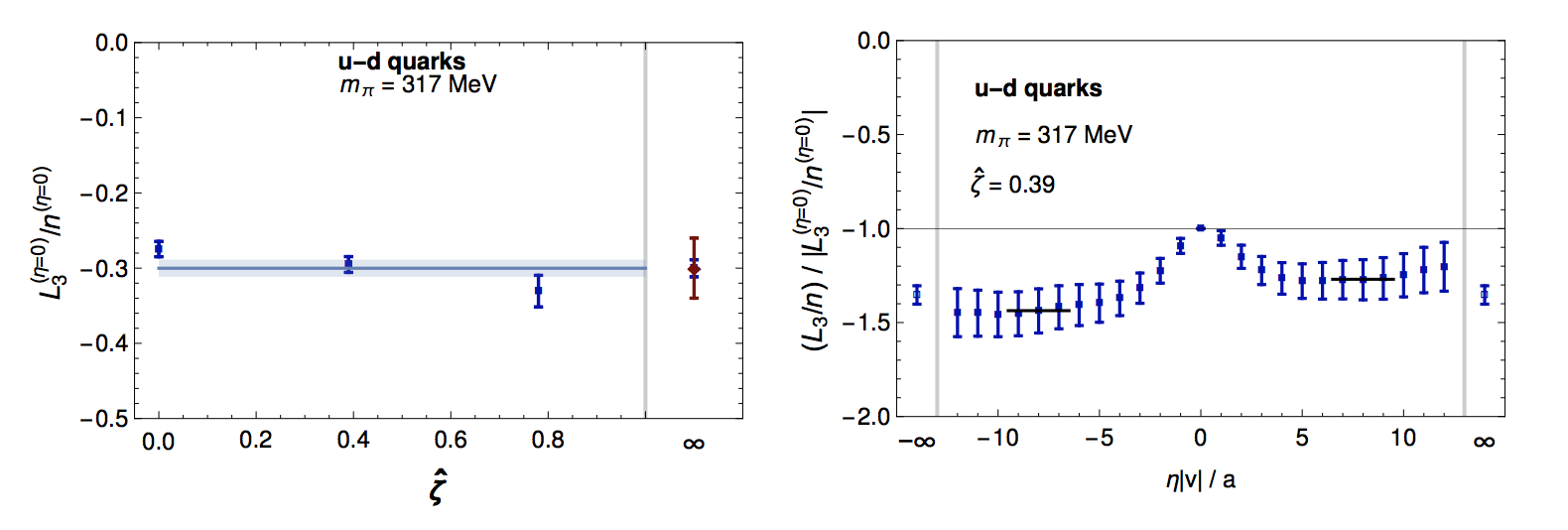}}
\caption{The preliminary results on the direct calculation of the quark iso-vector orbital angular momentum (OAM)~\cite{Engelhardt:2019lyy}. The left panel shows the consistency check for the direct calculation of the JI OAM and that from the difference between the Ji quark AM and quark helicity. The ratio of the OAM with the staple-shape link (becomes to JM OAM with proper perturbative matching at $\eta|v|\rightarrow \infty$ and infinite $\hat{\xi}\propto P_z$) and Ji OAM are presented in the right panel and hints that two definitions of OAM differs from unity.} \label{fig:T1+T2}
\end{figure}

\begin{figure}[htb]
 \centerline{ \includegraphics[width=1.0\hsize]{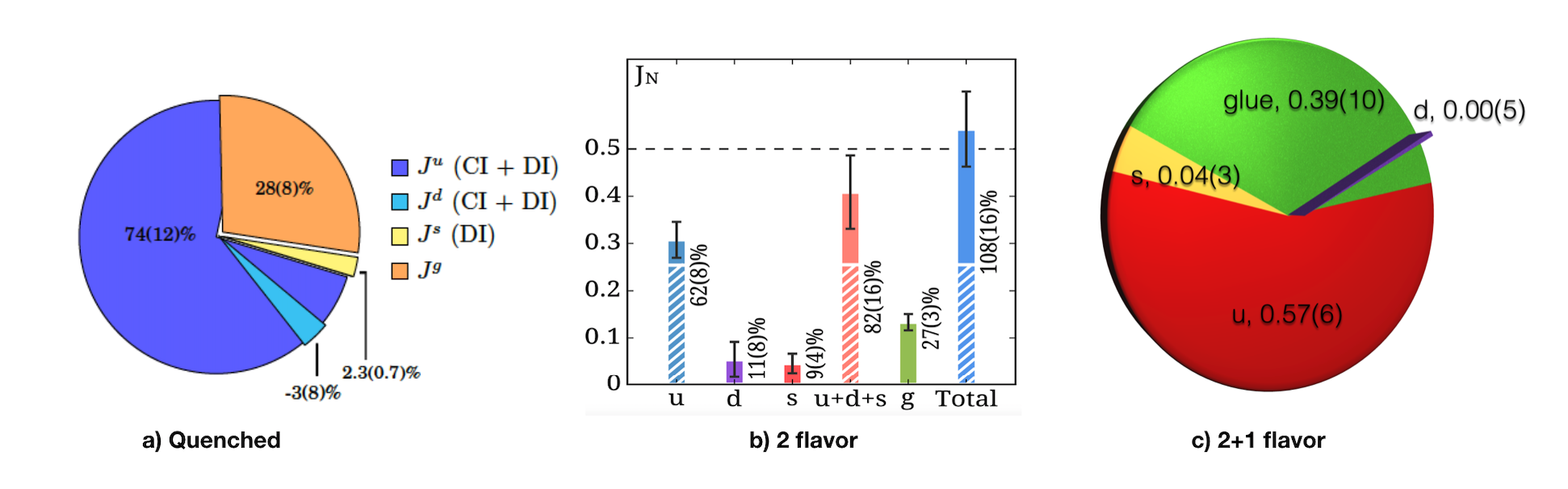}}
\caption{The angular momentum fractions from the quenched clover calculation (left panel)~\cite{Deka:2013zha}, 2 flavor Twisted-mass (+clover term) fermion calculation at physical pion mass (middle panel)~\cite{Alexandrou:2017oeh} and the preliminary non-perturbative renormalized 2+1 flavor overlap fermion on Domain Wall sea with 400 MeV pion mass (right panel). The first two results are consistent with each other while the last result prefer to a larger gluon angular momentum fraction with large statistical and uncontrolled systematic uncertainties (from the excited state contamination, chiral and continuum extrapolations).} \label{fig:AM}
\end{figure}

Recent preliminary progress on the calculation of quark OAM~\cite{Engelhardt:2019lyy}, shows that the direct calculation of the iso-vector Ji quark OAM can agree with the difference between the Ji quark AM and quark helicity (while the perturbative matching between the OAM ratio scheme and $\overline{\textrm{MS}}$ scheme and mixing with the quark helicity can be non-trivial and are still absent yet), as in the left panel of Fig.~\ref{fig:AM}. At the same time, the right panel of Fig.~\ref{fig:AM} shows the quark OAM with staple-shape link (becomes to JM OAM with proper perturbative matching at the $\eta|v|\rightarrow \infty$ and $\hat{\xi}\propto P_z\rightarrow\infty$ limit) is enhanced significantly compared to Ji OAM.

\begin{figure}[htb]
 \centerline{ \includegraphics[width=0.8\hsize]{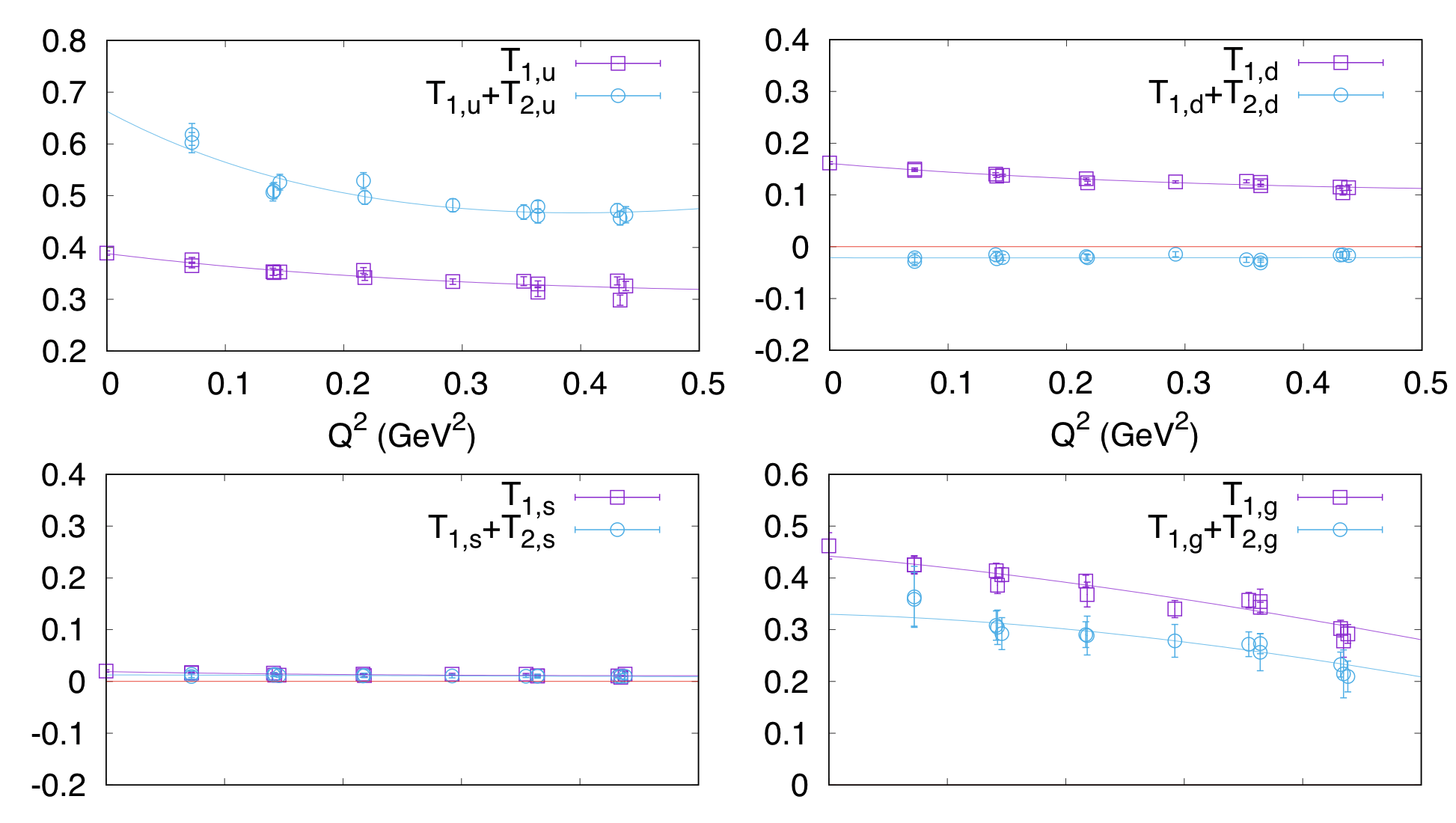}}
\caption{The preliminary results of the bare quark and glue EMT form factors as the function of $Q^2$, with 400 MeV pion mass overlap fermion on the 32ID ensemble~\cite{Blum:2014tka}, for $u$, $d$, $s$ quarks and gluon. The $T_1$ (red boxes) and $T_1+T_2$ (blue dots) form factors at $Q^2=0$ corresponds to the momentum and angular momentum fractions respectively. Both the contributions from the connected and disconnected insertions are included.} \label{fig:T1+T2}
\end{figure}

\begin{figure}[htb]
 \centerline{ \includegraphics[width=0.8\hsize]{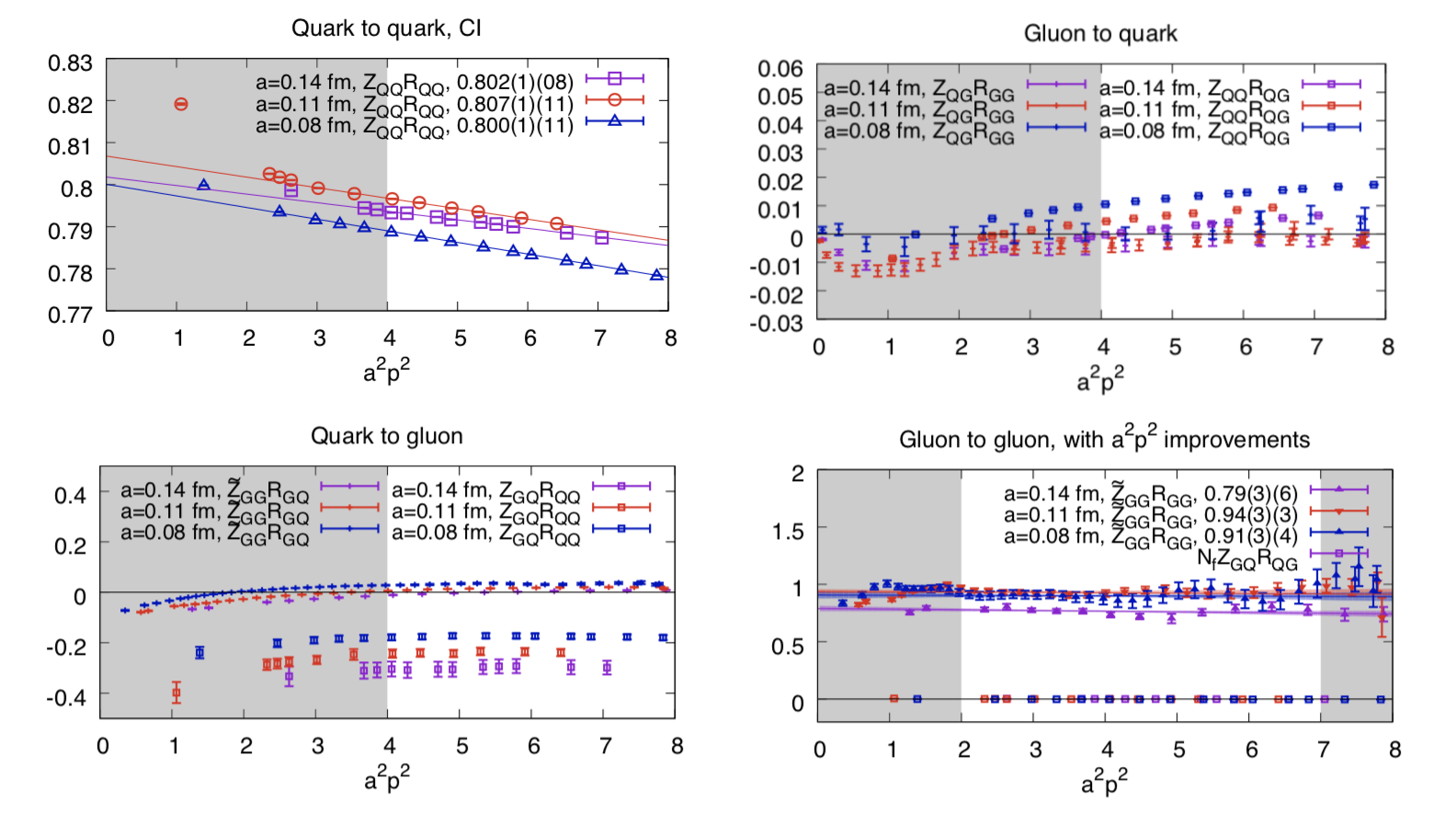}}
\caption{The renormalization and mixing of the quark and gluon energy momentum tensor (EMT) at $\overline{\textrm{MS}}$ 2 GeV as the function of $a^2p^2$~\cite{Yang:2018nqn}. The mixing from gluon to quark is small while that from quark to gluon can be large especially for the calculation on the 32ID enemsble at a=0.143 fm.} \label{fig:mix}
\end{figure}

To access the Ji quark and gluon AM, the usual way is calculating the EMT form factors of $T_1$ and $T_2$ defined by
\bea
\langle P'S' |\int d^3x \, {\mathcal T}^{\{0i\} q,g} | PS\rangle
  &=& \left(\frac{1}{2}\right) \bar{\chi}(P',S') \left[T_1(q^2)(\gamma^0\bar{p}^i 
   +  \gamma^i\bar{p}^0) 
   + \frac{1}{2m}T_2(q^2)\left(\bar{p}^0 (i \sigma^{i\alpha})
   +  \bar{p}^i (i \sigma^{0\alpha})\right) q_{\alpha}\right.\nonumber\\
  &+& \left.\frac{1}{m} T_3(q^2) q^0 q^i\right]^{q,g} \chi(P,S),
\eea
 at finite $Q^2$ first, and then extrapolate them to the forward limit $Q^2$, and then $T_1(0)$ and $T_1(0)+T_2(0)$ correspond to the momentum and AM fractions of the quark and gluon. Fig.~\ref{fig:T1+T2} shows the preliminary results of the bare $T_1$ and $T_1+T_2$ form factors and their z-expansion fits~\cite{Hill:2010yb} based on $\sim$ 1.0 fm source-sink separation, with 400 MeV pion mass overlap fermion on the 32ID ensemble~\cite{Blum:2014tka} using 170 MeV pion mass Domain Wall sea and DSDR gauge action at $a$=0.143 fm, including the contributions from both the connected and disconnect insertions. Four panels of Fig.~\ref{fig:T1+T2} are for the $u$, $d$, $s$ quarks and gluon respectively. It is obvious that $T_2$ form factor is positive for $u$ quark, negative in $d$ quark, and tends to be negative in the gluon case. 

Ref.~\cite{Yang:2018nqn} provided the non-perturbative normalization and renormalization of the quark and gluon symmetrized EMT including the mixing between them,
\bea
&&\left(\begin{array}{cc}
Z^{\overline{\textrm{MS}}}_{QQ}(\mu)+N_f\delta Z^{\overline{\textrm{MS}}}_{QQ} (\mu)&
N_fZ^{\overline{\textrm{MS}}}_{QG}(\mu)   \\
Z^{\overline{\textrm{MS}}}_{GQ}(\mu) &
Z^{\overline{\textrm{MS}}}_{GG}(\mu)
\end{array}\right)\equiv\left\{\left[\left(\begin{array}{cc}
Z_{QQ}(\mu_R)+N_f\delta Z_{QQ}&
N_fZ_{QG}(\mu_R)\\
Z_{GQ}(\mu_R)&
Z_{GG}(\mu_R)
\end{array}\right)\right.\right.\nonumber\\
&&\quad\quad\quad\quad\quad\quad\quad\quad\quad\quad\quad\quad\quad \left.\left.
\left(\begin{array}{cc}
R_{QQ}(\frac{\mu}{\mu_{R}})+{\cal O}(N_f\alpha_s^2) &
N_fR_{QG}(\frac{\mu}{\mu_{R}})\\
R_{GQ}(\frac{\mu}{\mu_{R}}) &
R_{GG}(\frac{\mu}{\mu_{R}})
\end{array}\right)
\right]|_{a^2\mu_R^2\rightarrow0}\right\}^{-1},
\eea
and $Z_{QQ}(\mu)=\left[\left(Z_{QQ}(\mu_R)R_{QQ}(\mu/\mu_R)\right)|_{a^2\mu_R^2\rightarrow 0}\right]^{-1}$. Note that the iso-vector matching coefficient $R_{QQ}(\frac{\mu}{\mu_{R}})$ has been obtained at the 3-loop level~\cite{Gracey:2003mr} while just the 1-loop level results of the other $R$'s are available~\cite{Yang:2016xsb}. 
Such a result can be applied there for the renormalization of the Ji AM. As in Fig.~\ref{fig:mix}, the mixing from quark to gluon can be large especially on the 32ID ensemble with a=0.143 fm. It enhances the Ji gluon AM much, and the renormalized Ji AM are plotted with the pie-chart in the right panel of Fig.~\ref{fig:AM} and the prediction are (with uncontrolled systematic uncertainties from the excited state contamination, chiral and continuum extrapolations): 2$J_u$=0.46(6), 2$J_d$=0.00(3), 2$J_s$=0.02(3), 2$J_g$=0.52(10). If we just use the value of $T_2(0)$ from the 32ID ensemble to correct the momentum fraction results in Ref.~\cite{Yang:2018nqn} which addressed the systematic uncertainties above, the prediction will be:
\bea
2J_u=0.57(6),\ 2J_d=0.00(5), 2J_s=0.04(3), 2J_g=0.39(10),
\eea  
and the quark OAM will be,
\bea
2L_u=-0.28(7),\ 2L_d=0.41(6),\ 2L_s=0.08(3).
\eea
Comparing to the previous quenched $\chi$QCD result~\cite{Deka:2013zha} (left panel of Fig.~\ref{fig:AM}) and 2 flavor ETMC result~\cite{Alexandrou:2017oeh} (middle panel) with 1-loop perturbative renormalization, the gluon AM is $\sim 1\sigma$ larger and the quark AM is smaller correspondingly. Further study on the systematic can provide a more accurate prediction of Ji AM.

\section{Summary} 

The Lattice QCD calculation of the proton spin, is an ongoing long story. Nowadays, the quark helicity calculation reaches some consensuses that $\Delta u \sim$ 0.8, $\Delta d \sim$ -0.4 which agree with the phenomenology determination based on the experiment, and $\Delta s\sim-0.05$ is just in the middle of the the result of Florian et al. ~\cite{deFlorian:2009vb} and COMPASS~\cite{Adolph:2015saz} with better precision. But the excited state contamination from the $\pi$N state can be crucial issue to examined carefully. For the gluon helicity, the pioneer investigation shows that the gluon spin under the Coulomb gauge is comparable with the present phenomenology fit of the experiment, while the result with a larger nucleon momentum $P_z$ and also the study on the other approaches are necessary to address the systematic uncertainties. All the present calculations on the Ji AM show that the contribution from both $d$ and $s$ quarks are small (but the OAM of the $d$ quark can be large), and some progress has been made in the matrix element calculation of the JM OAM. 

\bibliographystyle{unsrt}
\bibliography{reference.bib} 

\end{document}